\pdfoutput=1
\documentclass[aps,pre, twocolumn, groupedaddress]{revtex4-1} 
\usepackage{lipsum}
\usepackage{mathtools}
\usepackage{graphicx}
\usepackage{dcolumn}
\usepackage{amsmath}    
\usepackage{amssymb}
\usepackage{bm}
\usepackage{hyperref}
\usepackage{latexsym}
\usepackage{verbatim}
\usepackage{color}

\usepackage[caption=false]{subfig}

\def\beq{\begin{equation}}
\def\eeq{\end{equation}}

\def\beq{\begin{equation}}                          
\def\eeq{\end{equation}}                          
\def\bea{\begin{eqnarray}}                          
\def\eea{\end{eqnarray}}

\DeclareRobustCommand{\uvec}[1]{{%
  \ifcsname uvec#1\endcsname
     \csname uvec#1\endcsname
   \else
    \bm{\hat{\mathbf{#1}}}%
   \fi
}}

\preprint{}

\begin{document}


\title{Dynamics of a collection of active particles on a two-dimensional periodic undulated surface }
\author{Vivek Semwal}
\email{viveksemwal.rs.phy17@itbhu.ac.in}
\affiliation{Indian Institute of Technology (BHU) Varanasi, India 221005}
\author{Shambhavi Dikshit}
\email{shambhavidikshit.rs.phy18@itbhu.ac.in}
\affiliation{Indian Institute of Technology (BHU) Varanasi, India 221005}
\author{Shradha Mishra}
\email[]{smishra.phy@itbhu.ac.in}
\affiliation{Indian Institute of Technology (BHU) Varanasi, India 221005}
\date{\today}
\begin{abstract}
We study the dynamics of circular disk shaped active particles (AP) on a two dimensional periodic undulated surface. 
	Each particle has an internal energy mechanism which is modeled by an active friction force and it is  
	 controlled by an activity parameter $v_0$. It  acts as negative friction if the speed of the particle is smaller than
         $v_0$ and normal friction otherwise.
          Surface undulation is modeled
         by the periodic undulation of fixed amplitude and wavelength.
        The dynamics of the particle is studied for different activities and surface undulations (SU).
        Three types of particle dynamic is  observed on varying activity and SU:
	confined,  early time subdiffusion to diffusion and super diffusion to late time diffusion. 
	An effective equilibirum is established by showing the Green-Kubo relation  between the
        effective diffusivity  and the velocity auto-correlation function for 
	all  activities and small SU.

\end{abstract}

\maketitle

\section{Introduction}
In last few decades, active systems have become a subject of great interest \cite{sriramannulrev, marchettirmp, romanczuk, bechinger} due to  their unusual
properties in comparison to the system at thermal equilibrium.
Examples start from systems of  micron scale  like, bacterial colonies \cite{dell} up to a few meters like fish school \cite{fishschool}, bird flock \cite{flock} etc.
and also artificial microparticles like Janus particles \cite{filly, tailleur, cates2015, cates2013}.
Each constituent in these systems takes energy from their surroundings, and convert the energy into persistent motion
which leads to nonequilibrium behaviour.
Interestingly, the collective behaviour and phase separation is observed,  even in the absence of any external drive
\cite{sriramannulrev,marchettirmp,prithasoftmatter, tonertu5,tonertupre2018}.\\
A special class of active particles (AP), active Janus particles (AJP) are symmetrical in shape and hence do not have any alignment interaction \cite{filly, cates2013}.
One of the interesting features they exhibit is motility induced phase separation (MIPS) \cite{filly, redner,wittkowski,Palacci2013,prltwo2018,prr2020}.\\
AP also show interesting properties when kept in different environment \cite{Buttinoni2013}.
A recent study using  mesoscale hydrodynamic simulation found that the {\it E.Coli} bacteria can sense surface slip at the nanoscale
and hence can be used as {biosensor  \cite{biosensor1}.
Also, the study of  \cite{udit},} consider the motion of the chemically driven active colloid moving on the top
of two dimensional crystalline surface. It shows that the active colloid experiences competition between hindered
  and enhanced diffusion due to periodic surface and  activity respectively.
In other study by \cite{confiningpotential} it is reported that the  motion of the AP does not depend on the propulsion mechanism, but it is very
much influenced by the underlying surface properties.\\
A variety of theoretical and numerical studies are performed to
study the effect of a single and a collection of AP in different kinds of periodic, confined, 
and random medium or obstacles \cite{Palacci2013,Paoluzzi2014,biosensor,Pattanayak2019, Ray2014,Kmmel2013,maggi2016,Peruani2018}.
In some cases the presence
of periodic obstacles can produce directional transport \cite{Pattanayak2019, confinedmotion}, trapping \cite{sood, trapping}, and can be
used for sorting different kinds of AP \cite{bechinger}.  \\
Most of the theoretical understanding of AP is performed, where at each time step particle takes a constant step (self-propulsion speed). But in natural systems,  particle
 can have varrying self propulsion speed. It depends on its {\em activity}, inter-particle,  particle-medium interactions, and the thermal noise. \\
The origin of  {\em activity} can be due to an internal energy mechanism  \cite{energydepot}, and it is modeled through an active friction force.
The active friction force acts like negative friction and enhances the particle motion when it is moving slowly and
suppresses the motion when the dynamics become fast \cite{activefriction}. Previously it used to understand the dynamics of cells
in crowded environments and called as Schienbein Gruler (SG) friction \cite{udoerdmann}.\\

In this work, we study the dynamics of a collection of AP moving on a two-dimensional undulated surface with the active friction or SG friction.
The active friction is controlled by an activity parameter $v_0$.
For $v_0=0$, friction is like normal friction.
Surface undulation SU is controlled by a dimensionless parameter (SU) $\bar{h}$. The system is studied for
different activity and SU.
On the flat surface the dynamics of
particle is like a persistent random walk (PRW) \cite{randomwalk2}, and shows a
crossover from early time ballistic to late time diffusion.
Whereas on the  undulated surface, we find three distinct dynamics:
for small activity particle remains confined in one of the minima of the surface. For moderate activity,
particle remains stuck in a surface minimum for small time and randomly jumps from one minimum to another. Hence late time dynamics is diffusion with an
intermediate subdiffusion. For larger activity, waiting time in different minima is small and particle
shows the usual ballistic to diffusive motion.
The Green-Kubo relation
is found between the effective diffusivity   and velocity auto-correlation function (VACF) for the range of system parameters.\\
Our article is divided in the following sections: In  section \ref{secI},  we give the detailed description of our model. Section
\ref{secII} discusses about the results of numerical simulation of the system. In section \ref{secgk}, we establish a relation between
effective diffusivity calculated from the particles mean square displacement and VACF. 
In the last section \ref{secdis} we  conclude our result  and discuss about the future directions of our study.

\section{Model}\label{secI}
Our system consists of $N$ number of circular active  particles (AP) of radius $a_1$, moving on a two-dimensional substrate of dimension
$320 a_1 \times 320 a_1$. 
Substrate has periodic ups and downs of wavelength $l=10 a_1$.
Hence we call it {\it undulated} surface.  
Each particle on the surface, is defined by its position $\textbf{r}_i(t)$ and velocity
$\textbf{{v}}_{i}(t)$ at time $t$.
Activity of the particle is modeled by an active friction term which is controlled by an activity parameter $v_0$. Active friction 
arises  due to  an internal energy mechanism of the particle \cite{energydepot}.
It acts like a negative friction if the magnitude of particle velocity is smaller than $v_0$ and normal friction
otherwise. 
This type of friction is used  to model the dynamics
of cells in crowded enviournment \cite{franke, gruler, rapp, gruler2},  and it is called as  Schienbein and Gruler (SG) friction \cite{sg}. 
Particles also interact through a soft
repulsive interaction.
Hence, the equation of motion describing the dynamics of the particle involves (i) the active friction force,
(ii) soft repulsive interaction among the particles, (iii) the interaction between the particle and the  substrate and (iv) the thermal noise. 
Langevin's equation of motion governing the dynamics of the particle  is given by. 
\begin{equation}
	\frac{d\textbf{v}_i(t)}{dt}=\frac{1}{m}\bigg[-\gamma\bigg(1-\frac{v_0}{v_i}\bigg)\textbf{v}_i-  {\sum\limits_{j \neq i}{\bf F}_{ij}}-{
	{\boldsymbol{\mathcal{F}}}_i\bigg]+ { \sqrt{2D}}\boldsymbol{\xi}_{i} (t)}
\label{eqv}
\end{equation}
and the position is updated by
\begin{equation}
\frac{d\textbf{r}_i(t)}{dt}= \textbf {v}_i(t)
\label{eqr}
\end{equation}
here, the mass of the particle $m$ and friction coefficient $\gamma$  is taken as $1$.{ The ratio of the two defines the inertial time scale $\tau = (\gamma/m)$}.
The first term on the right hand side of equation  
\ref{eqv} is the active friction force, which acts like normal 
friction when magnitude of particle velocity $ {v}_i = \sqrt{v_{x_i}^2+v_{y_i}^2}$ $>$  $v_0$ and enhances the dynamics if $ {v}_i \textless v_0$. 
Hence { $\l_p = v_0\gamma/m$, is the persistent length or the run length, is the typical distance
travelled by the particle before it changes its velocity on the flat surface.} 
We  defined the dimensionless activity $\bar{v_0} = \frac{v_0 \gamma}{m a_1}$.
The second term, the force ${\bf F}_{ij}$ is the 
soft repulsive interaction among the particle. It is  obtained from the binary soft repulsive pair potential
$V(r_{ij}) = 
\frac{1}{2}k(r_{ij} - 2 a_1)^2$, where $r_{ij}=|{\bf r}_{j}-{\bf r}_{i}|$ is the distance between particle 
$i$ and $j$. 
The ratio of the strength of the interaction and the mass, $(k/m)^{-1/2}$ defined the elastic time scale.
The summation runs over all the particles. The force ${\bf F}_{ij}$ is non-zero if, $r_{ij}\leq 2 a_1$, else it is zero.
Further, the interaction force due to the  undulated surface  is given by { ${\boldsymbol{\mathcal{F}}} = -{\boldsymbol{\nabla} U(r_i)}$},
$U(r_i)= h sin(\frac{2\pi x_i}{l})sin(\frac{2 \pi y_i}{l})$
where ${\bf r}_i = (x_i, y_i)$ is the position of the $i^{th}$  particle 
on the flat surface. Although, surface has minima and maxima out of the plane, but we consider motion of the particle always in the plane and surface
is modeled  such that the speed of the particle increases (decreases) as it moves towards (away) to minima (maxima) and vice versa.
We define the dimensionless surface undulation, which is the ratio of surface interaction force with particle interaction force $\bar{h} = \frac{h m}{l^2 k}$.
The last term is the random thermal noise present due to medium. It is the Gaussian random force
with mean zero and correlation
{\begin{equation}
	<{\xi}_{il}(t){\xi}_{jm}(t')> = \delta_{lm}\delta_{ij}\delta \left (t-t' \right )
\end{equation}
$l$ and $m = 1,2$ are the indices for the coordinates in two-dimensions, $i$ and $j$ are the particle index.
$D$ is the strength of the noise \cite{noise}. 
If the system is in thermal equilibrium then $D$ can be fixed by the  temperature
of the medium. But no such constraint is imposed in active system and $D$ can be chosen as an independent parameter. In our present 
study, we fix $D = 0.045$ to keep the nosie term small. 
The control parameters in our model are  dimensionless activity $\bar{v_0}$ and  dimensionless surface interaction $\bar{h}$.
The inetraction among the particle is fixed  $k=1.0$. The 
characteristics of the system are studied for two independent parameters  $\bar{v_0}$ and $\bar{h}$, both changes  from $0$ to $10$ and $0$ to $1$ respectively.
We also compared the results for the two extreme limits of $k$, non-interacting $(k=0)$ and strongly interacting $(k=100)$, when the dimensionaless $\bar{h}$ becomes $>>1$ and $<<1$  respectively.
We also studied the  large friction $\gamma=100$ limit, when model can be 
mapped to overdamped motion of active Brownian particles.\\
We study the dynamics and the steady state of the particles moving on the surface, numerically integrating the two update 
Eqs. \ref{eqr} and \ref{eqv} using velocity Verlet algorithm \cite{velocityverlet, velocityverlet1} for the particle position and velocity. The numerical integration
is performed by choosing the time step $dt = 0.005$. We start with random initial positions and velocity directions  of 
all the particles. 
Once the update of above two equations is done for all  $N$ particles, it is counted as one simulation step. 
We perform the simulation for total simulation time up to $5 \times 10^6$.
All the physical quantities are calculated after waiting for the steady state time  upto 
$10^5$ and averaged over $20-50$ independent realisations. 
Simulation is performed for $N=11000$ active particles, hence packing fraction of particle density on the flat surface is  $\frac{N \pi a_1^{2}}{L^{2}}=0.31$.

{\section{Results}\label{secII}}
We first characterise the dynamics of particles  for different activities. 
Starting from the random positions and velocities, the 
particle dynamics is characterised by calculating the  mean square displacement (MSD), defined as
$\Delta (t) = \large \langle |{\bf r}(t+t_0)-{\bf r}(t_0)|^2\large \rangle$, where $<..>$,  implies average over many random 
initial conditions. $t_0 =1.0$ is the a fixed refrence time,
 the typical cross over time from ballistic to diffusive motion  on the flat surface for zero activity  $v_0=0.0$.
 The Fig. \ref{fig: 1}(a-d)  shows the plot of MSD, $\Delta (t)$ ${\it vs}.$ time $t$ 
for flat $\bar{h}=0$ and undulated surface $\bar{h} = 0.1$, $0.5$ and $1.0$ respectively. \\
We first describe the dynamics on the flat surface $\bar{h}=0.0$. The early time 
dynamics of particle is ballistic with $\Delta (t) \simeq t^2$ and as time progresses it shows a crossover to diffusion, $\Delta(t) \simeq t$.

\begin{figure} [hbt]
{\includegraphics[width=8 cm, height=6 cm]{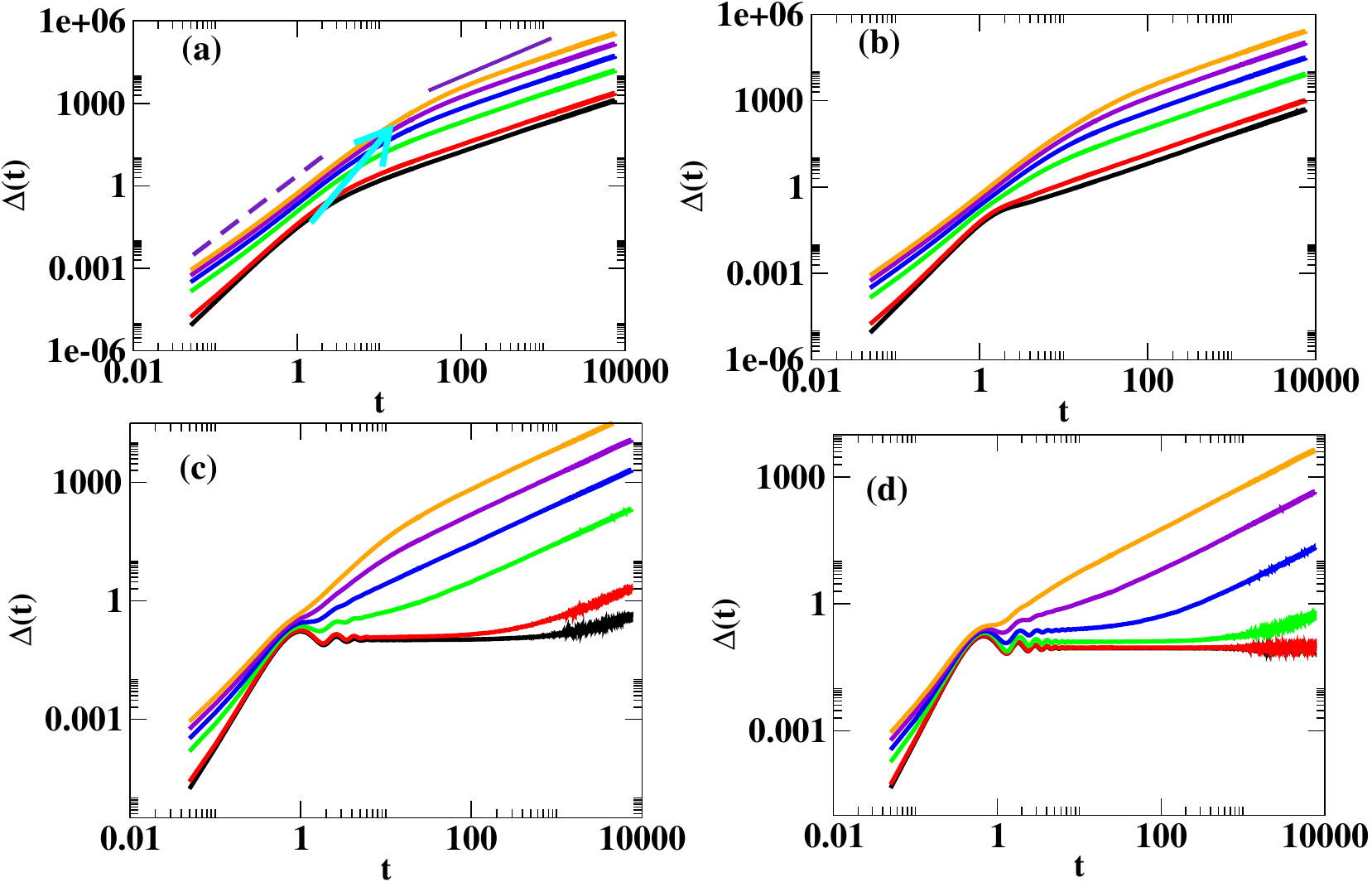}}
	\caption{(color online) Plot of MSD,  $\Delta(t)$ {\it vs.} time $t$ on $\log-\log$ scale, for
	$\bar{h}=0.0$(a), $\bar{h}=0.1$ (b), $\bar{h}=0.5$ (c),$\bar{h}=1.0$ (d),for the activity $\bar{v_0}$=( 0.0 black, 1.0 red,  4.0 green, 6.0 blue, 8.0 violet, 10.0 orange, lines respectively).
	The dashed and solid lines in (a) have slope 2 and 1 respectively.  The   arrow shows the increasing crossover time on increasing $\bar{v_0}$.} 
\label{fig: 1}
\end{figure}
The crossover time increases  on increasing $\bar{v_0}$. The  active nature of particle leads to enhanced persistent motion. 
Hence MSD can be compared with the result from persistent random walk (PRW) \cite{randomwalk}, where
$\Delta(t) = 2d{D_{eff}}t[1-\exp(\frac{-t}{t_c})]$,  where
$t_c$ is the  crossover time, $D_{eff}$ is the effective diffusivity and $d=2$ is the  dimensionality of space.
The $t_c$ and $D_{eff}$ obtained by fitting the data for MSD with PRW. 
When we turn on the SU, for  $\bar{v_0} \ge 8 $, dynamics  remains ballistic for small time and then it shows a smooth crossover 
to diffusive behaviour, as shown in Fig. \ref{fig: 1}(b). But as we increase SU  or for fixed SU decrease $\bar{v_0}$,
 MSD shows a  plateau for intermediate times, as shown in Fig.  \ref{fig: 1}(c-d). The extend of the plateau increases on increasing SU and 
decreasing $\bar{v_0}$ and for large $\bar{h} \gtrsim 0.8$ and small $\bar{v_0} \lesssim 1$, the extend of plateau present for very long time and particle is eventually confined.
In Fig.\ref{fig: 2} (a-b) we plot
the  scaled MSD, $\frac{\Delta(t)}{4D_{eff}t_c}$ {\em vs.} scaled time $\frac{t}{t_c}$. 
Data shows the  excellent scaling for the flat surface Fig. \ref{fig: 2}(a), which confirms that on the flat surface,  for all values of $\bar{v_0}$, the dynamics of particle is like PRW. 
As shown in Fig. \ref{fig: 2}(b),  motion on the undulated surface shows deviation from scaling, which is
 due to the transient arrest of particle in surface minima for small $\bar{v_0}$. The inset of Fig. \ref{fig: 2}(b) shows the zoomed plot of deviation from 
 scaling.
When two particles are stuck in the same surface minimum, then there is a competition between the activity and repulsion among the particles. 
and the both encourages the particles to come out. Hence the time spent in a surface minimum or length of the plateau increases on decreasing $\bar{v_0}$ and increasing $\bar{h}$. 
Interaction enhances the particle
dynamcis for a fixed activity $\bar{v_0}$. \\

We describe the particle dynamics in simple manner using real space snapshots of a single particle trajectory for fixed $\gamma=1.0$ and $k=1$, in Fig. \ref{fig: 3}(I-III). Fig. \ref{fig: 3}(b) 
shows the cartoon of part of surface. 
The dark  and bright  colors  show  the surface minima and maxima respectively.
\begin{figure} [hbt] 
	{\includegraphics[width=8 cm, height=3.5 cm]{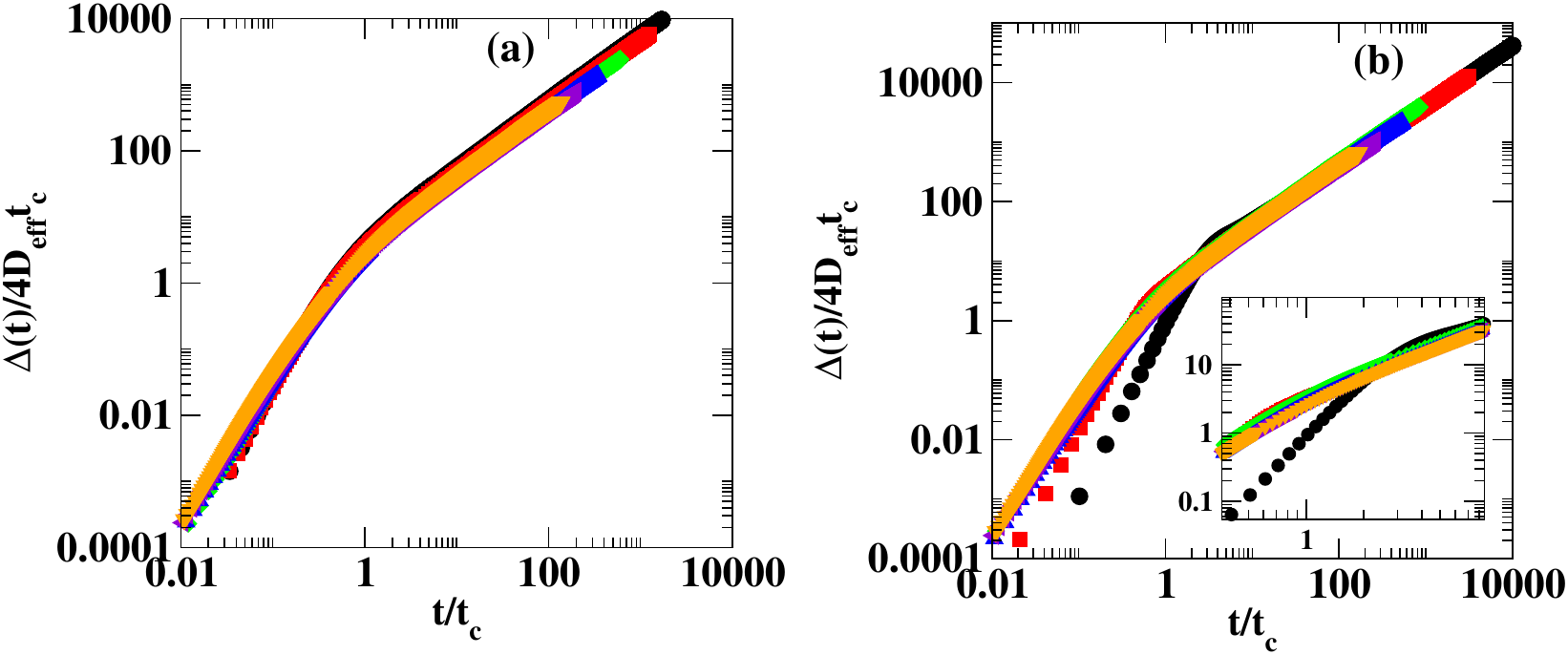}}
	\caption{(color online) Plot of the scaled MSD, $\frac{\Delta(t)}{4D_{eff}t_c}$ {\em vs.} scaled time $\frac{t}{t_c}$ on $\log-\log$ scale, for different $\bar{h}=0.0$ (a),$\bar{h}=0.1$ (b) for the same activities as in  Fig. \ref{fig: 1}(a). (b) (inset)  shows the zoomed plot near crossover time  $t_c$.}
	\label{fig: 2}
\end{figure}
For small values of $\bar{v_0} \lesssim 1.0$ and $\bar{h} \gtrsim 0.8$,
initially, (early time $\sim$ first few steps)  motion of particle is ballistic but soon it jumps into one of the minima and stays there (snapshot of particle
position for $\bar{h}=1.0, $ and $\bar{v_0} = 0.0$, as shown  in Fig. \ref{fig: 3}(I). Although, soft repulsive 
interaction among the particles will be maximum, when more than 
one particle sit in a minima but they do not come out  due to small activity.
Hence, MSD remains flat for the late time (as shown in Fig.{\ref{fig: 1}(d) (black circles). Increasing $\bar{v_0}$, leads to partial trapping of the particle in the 
minima and  particle starts moving from one minima to another after some transient time, as shown in snapshot Fig.\ref{fig: 3}(II) is for $\bar{h}=1.0$ and $\bar{v_0}=7.0$. So, after 
an intermediate time (plateau region) , MSD  starts growing linearly with time. 
Snapshot  in Fig. {\ref{fig: 3}(III) for $\bar{h}=1.0$ and 
$\bar{v_0}=10.0$. It shows the, particle's frequent jumps from
one minimum to another.\\

We further investigate the dynamics of particle by extracting the dynamic MSD exponent $\beta(t)$, defined by $\Delta(t) \sim t^{\beta(t)}$, hence $\beta(t)$ can 
be obtained by assuming MSD, $\Delta(t) \sim t^{\beta(t)}$. Hence $\Delta(10 t) \sim (10 t)^{\beta(t)}$, hence $\beta(t)$ can be obtained from 
the ratio of logarithmic (base 10) of two MSD's,
\begin{equation}
	\beta(t) = \frac{log_{10}[\Delta(10t)]}{log_{10}[\Delta(t)]}
\end{equation}
\begin{figure} [hbt]
        {\includegraphics[width=8cm, height=6cm]{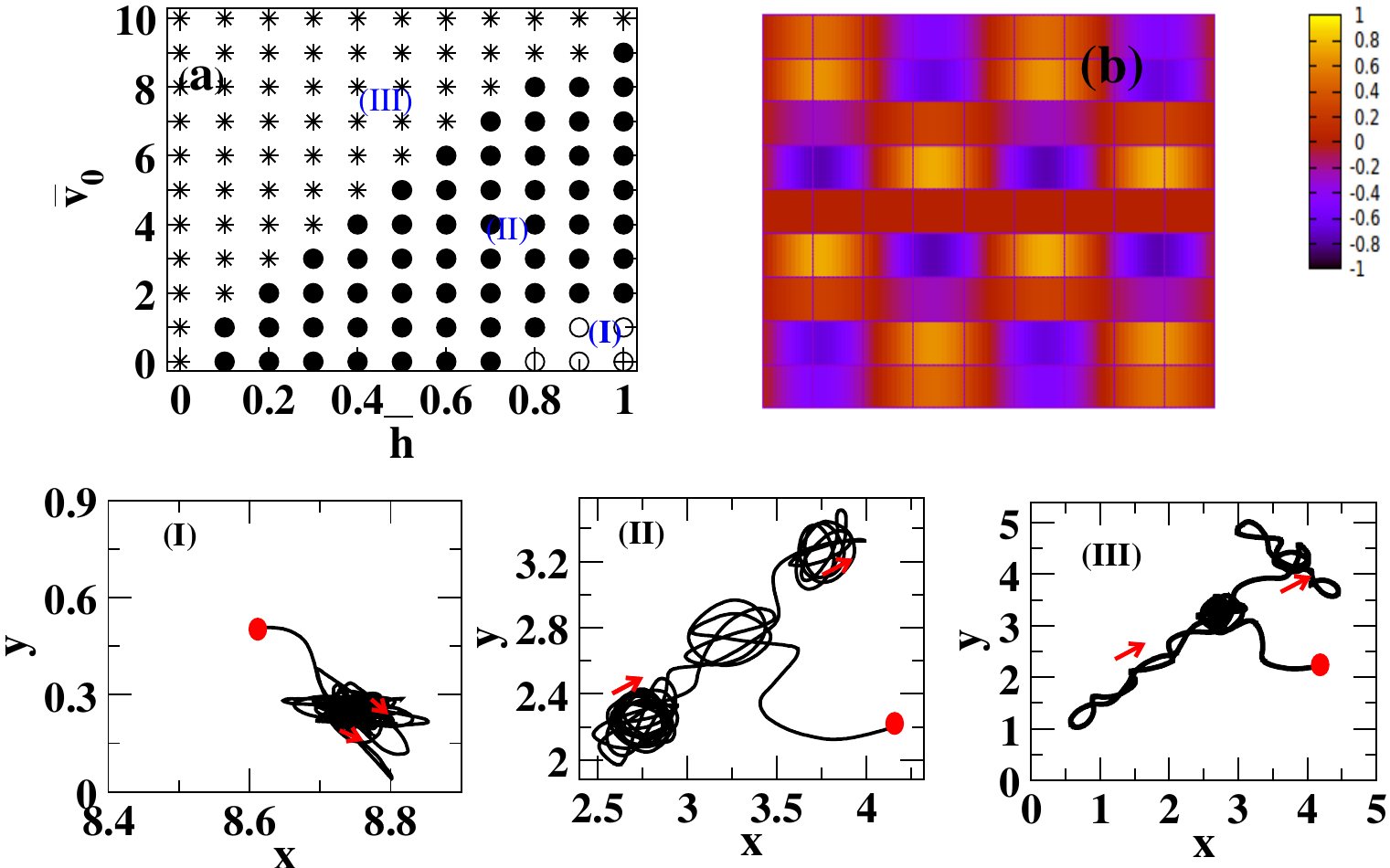}}
       \caption{(color online) (a) Phase diagram in the ($\bar{v_0}$, $\bar{h}$) plane,  region I (open circle), region II (filled circles) and region III (stars), three
        different phases (C), (SbD) and (SpD) respectively on the linear-linear scale. (b) Cartoon of a part of  the surface. The bright blue color  and bright yellow color show the location of minima and maxima respectively and color bar shows the height of the surface.
         (I-III) shows the trajectory of a single particle for parameters $(\bar{v_0}, \bar{h})$ = $( 0.0,1.0 ), (7 ,1.0 ), (10.0 ,1.0 )$ 
	where particle is in three different regions (I)-(III) respectively. For all (I-III) ($t=6000$) time duration of particle trajectory is same and red dot
        is the location of particle at the start of the trajectory, {arrow shows the direction of trajectory after every 2000 times.}}
        \label{fig: 3}
\end{figure}
Fig. \ref{fig: 4}(a-d) shows  the  plot of $\beta(t)$ {\em vs.} $t$ for flat  and undulated surfaces,  $\bar{h}$= $0, 0.1, 0.2$ and $1.0$ respectively.
For all  $\bar{v_0}$, late time value of $\beta$ either  $0$ (confinement) or $1$ (diffusion).
Approach to the late time dynamic, depends upon the  SU and activity. On the flat surface  $\bar{h}=0$, for all activity and greatest $\bar{h} = 1.0$, for large $\bar{v_0} \ge 10.0$,  
approach is always through an early time 
superdiffusion $\beta>1$ to late time diffusion $\beta=1$, but  for moderate $\bar{v_0} <10.0$, 
approach to $\beta=1$ is through an intermediate 
subdiffusive regime, where $\beta<1$. 
Also for very small activity and large SU ($\bar{v_0} \lesssim 1$ and $\bar{h} \gtrsim 0.8$), motion is  confined in one of the surface
minimum. 
Hence, the dynamics of particle is of three types: (i) late time confinement $\beta(t)=0$ (C), 
(ii) approach to diffusion $\beta(t)=1$ from intermediate subdiffusion $\beta(t)<1$ (SbD) and (iii) Initial superdiffusion $\beta(t) >1$ to late time 
diffusion $\beta(t)=1$ (SpD).
Hence for sufficiently large activity $\bar{v_0} \gtrsim 1$, the asymptotic  dynamics of particle moving 
on undulated surface is always diffusive, only route to the steady state is different. \\
We also compared the results for large friction coefficient, when model can be compared with the overdamped dynamics of  ABP  
on undulated surface. We find much slower dynamics for
the large friction limit {$\gamma=100$} as shown in \ref{fig: 5}(a-b). We also  compared the results for non-interacting $k=0$ and large interaction $k=100$ in Fig. \ref{fig: 5}(a), where
dynamics can be similar to particle moving on strong surface and flat surface respectively.  Hence effective dynamics becomes slower and enhanced for the two extreme cases as shown in 
Fig. \ref{fig: 5}(b). \\
Further, we propose that, the  underlying surface acts like a 
medium with an effective temperature, in which particles are moving. To confirm this we  compare the effective diffusivities from MSD
with the velocity auto-correlation function  VACF, $C(t) =  <{\bf v}_i(t+t_0)\cdot {\bf v}_i(t_0)>$ using Green-Kubo (GK) relation \cite{gk, gk1}. 
To our surprise we find that for flat as well as moderate $\bar{h} \lesssim 0.2$, the GK relation is satisfied for all values of $\bar{v_0} \in (0,10.0)$. Details of our study we discuss next.

\begin{figure}[hbt]
        {\includegraphics[width=8cm, height=6 cm] {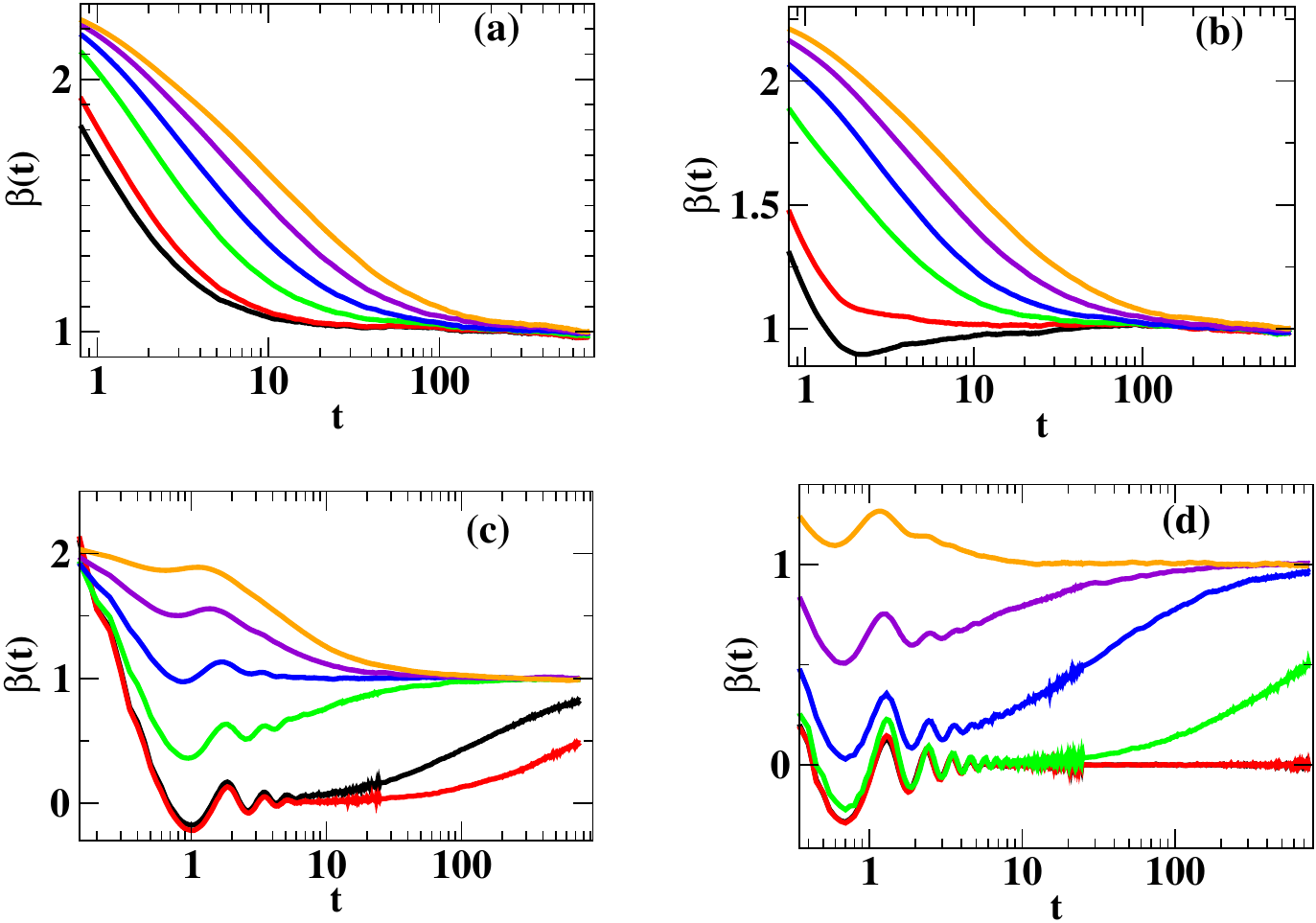}}
	\caption{(color online) (a-d) Plot of the  dynamic MSD exponent $\beta(t)$ {\it vs.} time $t$ on $\log-linear$ scale,  for different $\bar{h}$ and $\bar{v_0}$ same as in Fig. \ref{fig: 1}(a).}
	\label{fig: 4} 
\end{figure}
\begin{figure}[hbt]
        {\includegraphics[width=9cm, height=3.5 cm] {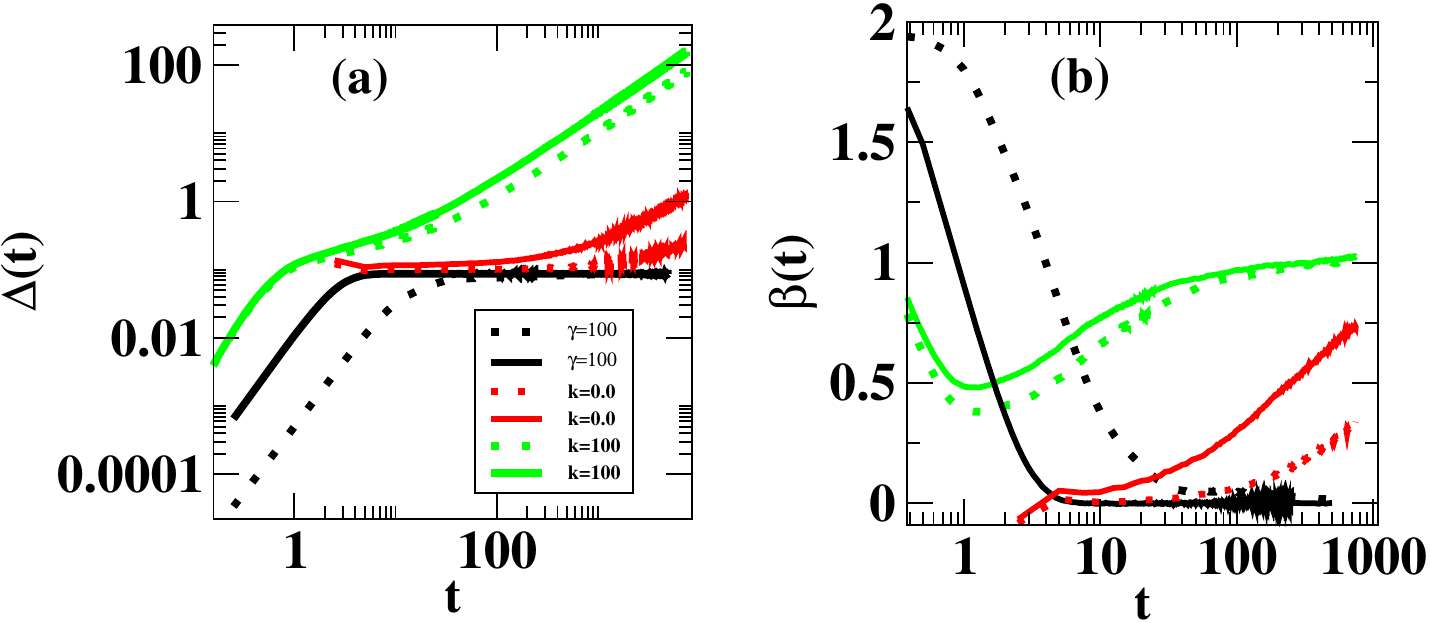}}
	\caption{(color online) Plot of the MSD (a), $\beta(t)$ {\it vs.} time $t$ on $\log-linear$ scale for plot (a) in (b),  different value of $k$ and $\gamma$, dotted and solid line repectively for $\bar{v_0}=0.0,1.0$ .}
	\label{fig: 5} 
\end{figure}

\subsection {Green-Kubo relation}\label{secgk}
\begin{table*}[ht] 
  \caption{Comparison of diffusivity from MSD and VACF} 
\label{table:vivek}     
\begin{tabular}{ |p{1cm}|p{2.4cm}|p{2cm}|p{2.4cm}|p{2cm}|p{2.4cm}|p{2cm}|}
\hline
 
$\bar{v_0}$&$\Delta D_{eff}(\bar{h}=0.0)$&$\Delta \mathcal{D}(\bar{h}=0.0)$&$\Delta D_{eff}(\bar{h}=0.1)$&$\Delta \mathcal{D}(\bar{h}=0.1)$&$\Delta D_{eff}(\bar{h}=0.2)$&$\Delta \mathcal{D}(\bar{h}=0.2)$ \\ \hline
 1&0.811$\pm$0.002 & 0.82$\pm$0.0046&1.12$\pm$0.006&1.09$\pm$0.003&1.90$\pm$0.006&1.86$\pm$0.006\\
 \hline
 2 &2.25$\pm$0.008 & 2.45$\pm$0.008& 3.37$\pm$0.0028&3.63$\pm$0.015&6.41$\pm$0.006&5.95$\pm$0.009\\
 \hline
 3&5.27$\pm$0.019 & 5.51$\pm$0.013& 7.87$\pm$0.008&7.90$\pm$0.01&17.06$\pm$0.001&15.08$\pm$0.004\\
 \hline
4&11.09$\pm$0.03&11.41$\pm$0.028&14$\pm$0.015&16.27$\pm$0.017&37.7$\pm$0.004&34.65$\pm$0.01\\
 \hline
 6&36.69$\pm$0.013 &40$\pm$0.064   &63.37$\pm$0.05&63.18$\pm$0.02&169.96$\pm$0.019&154.21$\pm$0.041\\
 \hline
 8&113.18$\pm$0.09&113.51$\pm$0.065&210.25$\pm$0.25&206.36$\pm$0.38&650.61$\pm$0.12&593.78$\pm$0.05\\
 \hline
 10&266.90$\pm$0.34 &265.31$\pm$0.39&551$\pm$0.19&530$\pm$0.26&2011.9$\pm$0.25&1857.26$\pm$0.2\\
 \hline
\end{tabular}\\
\end{table*}
 We  first measure the  VACF  for the flat and different SUs. The VACF decays exponentially to zero  on the flat surface, 
 $C(t)=C_0(\exp(-t/\tau)$, where $C_0$ is the correlation for $t=0$ and $\tau$ is the decay  time and it increases with increasing activity. 
 On undulated surface, after the initial exponential decay, the VACF oscillates about zero. The oscillations are due to periodic 
 trapping and untrapping of particle due to finite depth of the surface. We estimate the decay time $\tau$ from the exponential decay.
 In  Fig. \ref{fig: 6} (a-b) we
  plot the  scaled VACF, $C(t)/C_0$ {\it vs.} scaled time $t/\tau$ for flat surface and for $\bar{h}= 0.1$. 
 Data shows nice scaling collapse for all $\bar{v_0}$ on the flat surface. But on the undulated surface,  it  deviates at late times,
 due to periodic oscillations in VACF.
  Now we assume an effective equilibrium and calculate  the diffusivity $\mathcal{D} (\bar{v_0}, \bar{h})$ by the late time  
  $\lim_{t \rightarrow \infty} \frac{\int_{0}^{t}(<v(t')v(0)>)dt'}{d}$, 
  where $d$ is dimensionality of the space.
   We further compare it with 
  the effective  diffusivity calculated from MSD, $D_{eff}(\bar{v_0}, \bar{h}) = lim_{t \rightarrow \infty} \frac{\Delta(t)}{2 d t}$.


\begin{figure} [hbt]
	{\includegraphics[width=8 cm, height=3.5 cm]{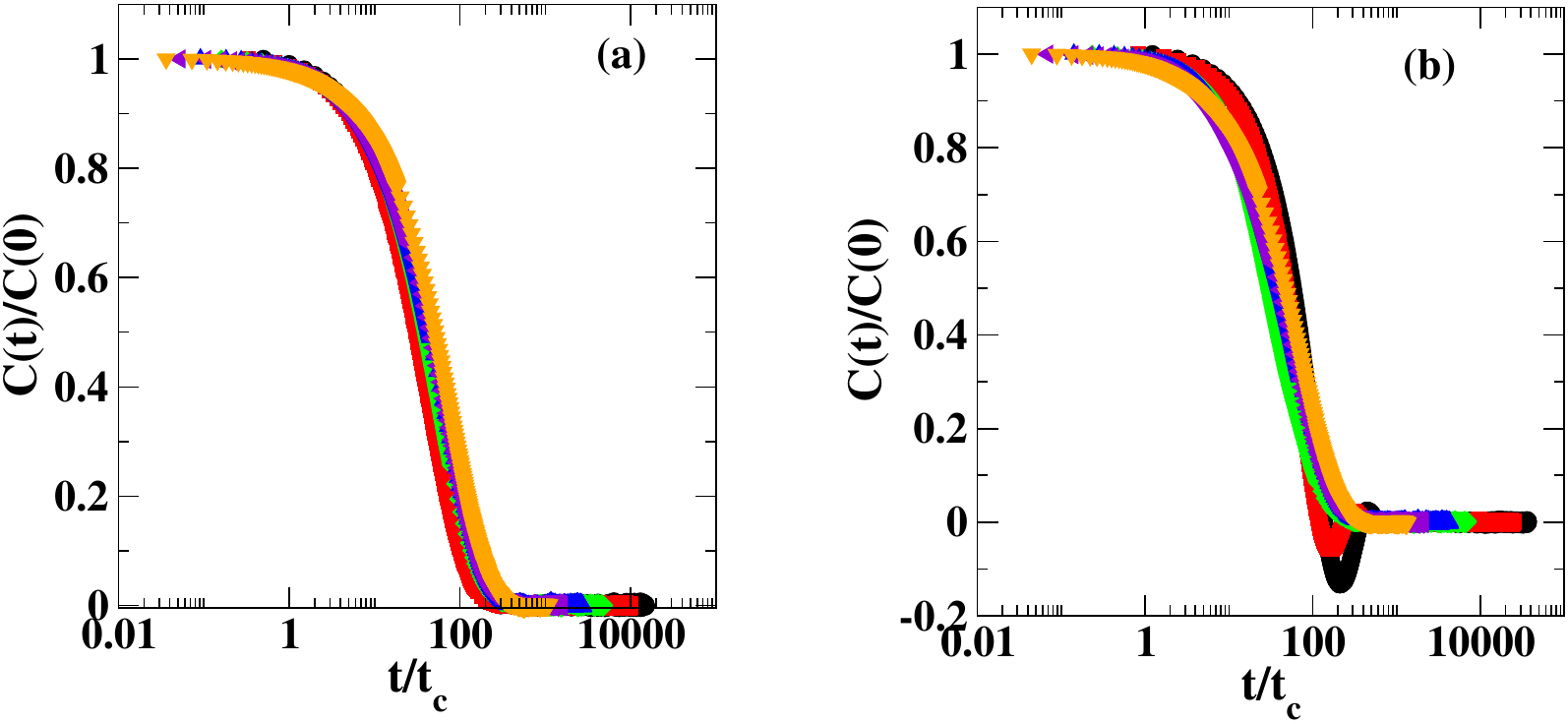}}
	\caption{(color on line)  Plot of the scaled VACF  for (a) $\bar{h}=0.0$ and (b) $\bar{h}=0.1$,  on the log-linear scale. The other parameters are same as in Fig. \ref{fig: 2}.}
	\label{fig: 6}
\end{figure}

 
\begin{figure}
	{\includegraphics[width=8.0cm, height=4.0 cm]{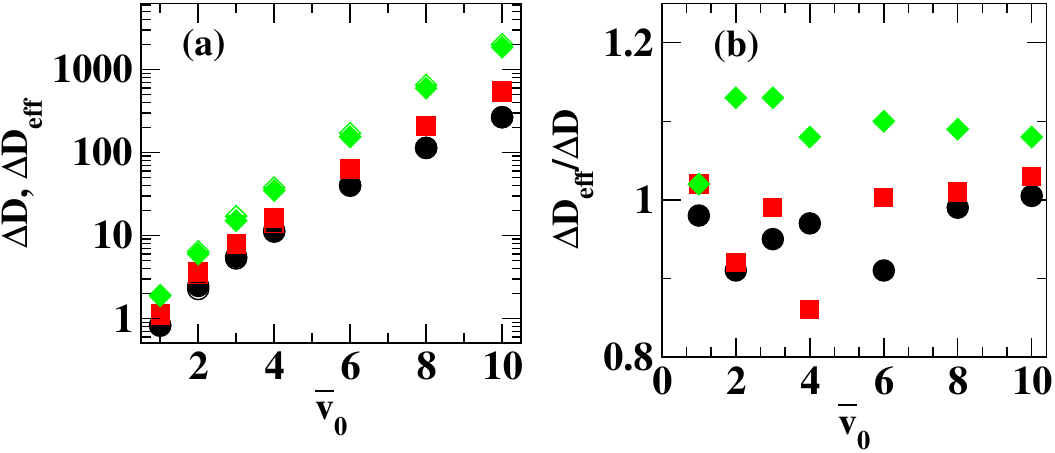}}
	\caption{(color online){ Comparison plot of the two relative diffusivities from MSD and VACF, $\Delta D_{eff}$ and $\Delta \mathcal{D}$ respectively with respect to $\bar{v_0}$.  Open and filled
	symbols are for $\Delta D_{eff}$ and $\Delta \mathcal{D}$ respectively. The three symbols (circles, squares and diamonds) are for three different  $\bar{h}=0.0$ and $\bar{h} = 0.1$ and $\bar{h}=0.2$ respectively. Plot (b) shows $\Delta D_{eff}$/$\Delta \mathcal{D}$ ratio of diffusivities  from MSD and VACF. Error bars are of the order of symbol size.}}
	\label{fig: 8}
\end{figure} 
 Fig.  \ref{fig: 8} shows the  plot of the comparison between the two relative diffusivities  from MSD and VACF on the linear-log scale, 
$ \Delta D_{eff} = \frac{(D_{eff}(\bar{h},\bar{ v_0})-D_{eff}(\bar{h}, 0))}{D_{eff}(\bar{h}, 0)}$ and $ \Delta \mathcal{D} = \frac{(\mathcal{D}(\bar{h},\bar{ v_0})-\mathcal{D}(\bar{h}, 0))}{\mathcal{D}(\bar{h}, 0)}$  
 {\it vs.} $\bar{v_0}$ respectively, for three different $\bar{h} = 0, 0.1$ and $0.2$. The $D_{eff}(\bar{h}, 0)/\mathcal{D}(\bar{h}, 0)$ is the diffusivity for zero $v_0$ or for passive system. 
 It is larger on the flat surface and decreases on increasing $\bar{h}$. In the table \ref{table:vivek} we list
 the two relative diffusivities for flat $\bar{h}=0.0$ and undulated surface $\bar{h}=0.1$ and $0.2$. 
 On the flat surface the two relative diffusivities shows good match and hence GK relation is satisfied.
 On the undulated surface, for smaller $\bar{v_0}$, data shows good match  for all $\bar{v_0}$ and $\bar{h}$. Hence for small $\bar{v_0}$ and 
 $\bar{h}$ an effective equilibrium is found in this nonequilibirum system. Fig. {\ref{fig: 8}(b) shows the comparison (ratio) plot of the two relative diffusivities form MSD and VACF. {As it is very clear for
all $\bar{v_0}$, data for ratio fluctuates around 1, and approach to 1, for larger $\bar{v_0}$.
Any deviation we find is due to partial trapping of particle in surface. For small $\bar{v_0}$, time
spend in trapped state or plateau is longer hence more deviation from GK relation.}}
Hence in such active system an effective equilibrium can be establsihed with respect to relative diffusivity as in corresponding  passive system. 

{\section{Discussion}\label{secdis}}
We have studied the dynamics and steady state of a collection of AP moving on a two-dimensional periodically undulated surface.
The activity of the particle is present due to an internal energy mechanism, which introduces an active friction \cite{sg}, which
enhances the particle motion when it slows down and suppresses the motion when it tries to accelerate. The activity $\bar{v_0}$ and
SU $\bar{h}$ are the two control parameters of the system. On the flat surface, $\bar{h}=0$, dynamics of the particle is like PRW with initial ballistic to late
time crossover to diffusion.
The crossover time increases by increasing $\bar{v_0}$. On the undulated surface we find a  systematic deviation from PRW and
particle shows the transient arrest in different surface minima. Due to this, the MSD shows a plateau for small $\bar{v_0}$ and larger $\bar{h}$.
The particle shows the three types of motion: (i) confined (C), (ii) from initial subdiffusion to late time diffusion (SbD)
and (iii) initial superdiffusion to late time diffusion  (SpD). We draw a phase diagram in the plane of ($\bar{v_0}$, $\bar{h}$).
Hence final state and route to the late time dynamics of the particle very much depend on its activity and surface
characteristics. \\
Although the system is highly nonequilibrium, we find that for moderate $\bar{h} \lesssim 0.2$  the
Green-Kubo relation is satisfied between the effective diffusivity and velocity auto-correlation function. \\
Our study provides a phase diagram for AP moving under active friction. For finite activity the late time dynamics is diffusive,
but route to diffusion is different and depends on surface characteristics and activity. Whereas on the flat surface 
motion is always like PRW. Hence our study provide the characteristics of AP moving on periodic surface.
Our work shows that different types of motion can be generated by tuning the surface and particle interaction. Hence these  results
can be used for various technological and pharmaceutical applications of AP.\\
The current study is  limited for the periodic surface, it will be interesting to
find the behaviour of particles on the surface with random maxima and minima which is  present in many biological systems \cite{biological}. 

{\bf{Acknowledgement}}\\
We thank Paramshivay  supercomputing center 
facility at I.I.T.(BHU) Varanasi. VS thanks DST INSPIRE(INDIA) for the research fellowship. S. M. thanks DST,SERB(INDIA),project no. ECR/2017/000659 for partial financial support.\\
{\bf{Author contribution statement }}
VS and SM designed the project. VS and SM developed the numerical code, and VS executed it. VS and SM contributed equally. VS, SM and SD contributed in analysing the result and preparing the manuscript. 

\end{document}